\documentclass[twocolumn]{aastex63}

\usepackage{graphicx,array,multirow}	
\usepackage{amsmath}	
\usepackage{amssymb}	
\usepackage{bm}		

\usepackage[T1]{fontenc}
\usepackage{ae,aecompl}

\usepackage{newtxtext,newtxmath}
\usepackage{xcolor}
\usepackage{pifont} 
\newcommand{\cmark}{{\color{green} \ding{51}}}%
\newcommand{\xmark}{{\color{red} \ding{55}}}%

\submitjournal{ApJ}

\shorttitle{Merger rates, kpc-scale wandering and supernova feedback}
\shortauthors{Barausse et al.}
\graphicspath{{./}{figures/}}

\begin{document}

\title{Massive black hole merger rates: the effect of kpc separation wandering and supernova feedback}

\correspondingauthor{Enrico Barausse}
\email{barausse@sissa.it}

\author{Enrico Barausse}
\affiliation{SISSA, Via Bonomea 265, 34136 Trieste, Italy and INFN Sezione di Trieste}
\affiliation{IFPU - Institute for Fundamental Physics of the Universe, Via Beirut 2, 34014 Trieste, Italy}
\affiliation{Institut d'Astrophysique de Paris, CNRS \& Sorbonne Universit\'e, UMR 7095, 98 bis bd Arago, 75014 Paris, France}

\author{Irina Dvorkin}
\affiliation{Institut d'Astrophysique de Paris, CNRS \& Sorbonne Universit\'e, UMR 7095, 98 bis bd Arago, 75014 Paris, France}

\author{Michael Tremmel}
\affiliation{Yale Center for Astronomy and Astrophysics, Physics Department, P.O. Box 208120, New Haven, CT 06520, USA}

\author{Marta Volonteri}
\affiliation{Institut d'Astrophysique de Paris, CNRS \& Sorbonne Universit\'e, UMR 7095, 98 bis bd Arago, 75014 Paris, France}

\author{Matteo Bonetti}
\affiliation{Dipartimento di Fisica ``G. Occhialini'', Universit\`a degli Studi di Milano-Bicocca, Piazza della Scienza 3, 20126 Milano, Italy}










\begin{abstract}
       We revisit the predictions for the merger rate of massive black hole binaries detectable by the Laser Interferometer Space Antenna (LISA) and their background signal for pulsar-timing arrays. We focus on the effect of the delays between the merger of galaxies and the final coalescence of black hole binaries, and on supernova feedback on the black hole growth. By utilizing a semi-analytic galaxy formation model, not only do we account for the driving the evolution of binaries at separations $\lesssim 1$ pc (gas-driven migration, stellar hardening and triple/quadruple massive black hole systems), but we also improve on previous studies by accounting for the time spent by black hole pairs from kpc down to pc separation. We also include the effect of supernova feedback, which may eject gas from the nuclear region of low-mass galaxies, thus hampering the growth of black holes via accretion and suppressing their orbital migration in circumbinary disks. Despite including these novel physical effects, we predict that the LISA detection rate should still be $\gtrsim 2 \mbox{yr}^{-1}$, irrespective of the model for the black hole seeds at high redshifts. Scenarios where black holes form from $\sim100 M_\odot$ seeds are more significantly impacted by supernova feedback. We also find that for  detectable events, the merging black holes typically have mass ratios between $\sim 0.1$ and $1$. Predictions for the stochastic background in the band of pulsar-timing array experiments are instead rather robust, and show only a mild dependence on the model.
\end{abstract}

\keywords{black hole physics -- galaxy dynamics -- gravitation -- gravitational waves}


\section{Introduction} \label{sec:intro}

The origins of massive black holes (MBHs) and the nature of their co-evolution with their host galaxies remain fundamental questions in astrophysics, which current and future gravitational-wave detectors may help decipher. MBHs are ubiquitous in massive galaxies \citep{Gehren84, Kormendy1995} as well as in a fraction of low-mass dwarf galaxies \citep{reines11, reines13, Baldassare2019}. Feedback from active galactic nuclei (AGNs), powered by growing MBHs, is commonly thought to regulate and quench star formation in massive galaxies \citep{Croton2006}, and possibly in dwarf galaxies as well \citep{Dickey2019, Sharma2019}. Despite their importance to galaxy formation theory, the mechanisms that drive and regulate MBH growth in galaxies, as well as the physical processes surrounding AGN feedback, are still not well understood. Scaling relations between MBH mass and galaxy properties are indicative of coeval growth \citep{Kormendy2013, Mcconnell2013, Schramm2013}, although as the census of MBHs in the local universe improves, such relationships are found to be more complicated than previously thought \citep{VolonteriReines16, Shankar2016, Barausse2017, greene19}. 

An important limitation of our understanding of MBHs and their effect on galaxies is related to the difficulty of studying their early evolution at high redshift, since only the most luminous, rapidly growing MBHs are accessible by observations.
Gravitational waves represent an intriguing window into the history of MBHs,
because their propagation through the universe, unlike that of electromagnetic radiation, is essentially unobstructed.
The future Laser Interferometer Space Antenna \cite[LISA;][]{LISA2017} will be able to detect gravitational waves emitted by merging MBHs with masses $10^4$--$10^7 M_{\odot}$ out to redshifts greater than 20, while providing accurate estimates on MBH mass, spin, and MBH binary orbital parameters \citep{Klein2016,colpi19}. Such detections promise new insight into the MBH population at early times, and may place unique constraints on models of MBH formation and growth \citep{Sesana2007b, Volonteri2009, Berti2008, Sesana2011a, Plowman2011, Klein2016, Ricarte2018b, 2019MNRAS.486.4044B}. 

Currently, pulsar-timing arrays are searching for (unresolved) gravitational-wave signals from lower redshift ($z<2$), higher mass ($> 10^8 M_{\odot}$) MBH binaries using Milky Way pulsars~\citep{Lentati2015,Arzoumanian2016,Shannon2015,Verbiest2016,Arzoumanian2018,Perera2019}. While the sensitivity of pulsar-timing arrays is still increasing with time, the absence of a detection so far, and the
resulting upper limits on the background of unresolved gravitational 
waves at nHz frequencies, are already placing
significant constraints on models of
 MBH mergers~\citep{Wyithe2003,Sesana_Vecchio2008,McWilliams2014,Rajagopal1995,Jaffe2003,Sesana2013,Ravi2015,Sesana2016,Sesana2009,Ravi2012,Kulier2015,Kelley2017,2019MNRAS.486.4044B}, even ruling out the most extreme ones~\citep{McWilliams2014}.
 However, it is critical to note  that much of the astrophysics that can be inferred from gravitational-wave detectors relies heavily on the models used to predict the MBH binary and merger populations and their various underpinning assumptions.

Semi-analytic models of galaxy and MBH formation and evolution are powerful tools for deriving astrophysics from gravitational-wave detections, as well as informing the experimental setups themselves \citep[e.g.][]{Sesana2004,Volonteri2008, Volonteri2009,Barausse2012,Klein2016,Ricarte2018a, 2019MNRAS.486.4044B}. These models relate the hierarchical formation of dark matter halos to the evolution of galaxies and their MBHs. Semi-analytic models have been successful in reproducing the observed evolution of galaxy morphology, color, star-formation rate and luminosity functions and, because they are relatively inexpensive, have been used to explore the wide parameter space of galaxy formation physics \citep[e.g.][]{Somerville1999,Somerville2008, Croton2006}. Critically, modern semi-analytic galaxy formation models include prescriptions for MBH growth and feedback, similarly constrained by both galaxy observations and quasar luminosity functions \citep{Somerville2008, Barausse2012, Sesana2014, Ricarte2018a}. By tracking the properties of merging galaxies and the formation of MBH binaries over cosmic time, these simulations can predict how gravitational-wave signals relevant to both LISA and pulsar-timing arrays are affected by various physical processes, such as feedback and MBH dynamical evolution on various scales.

The process of forming an MBH binary begins when two galaxies embedded in their dark matter halos merge, and the MBHs are still separated by 10s-100s kpc.
The MBH orbital evolution then proceeds down to separations of a few hundred pc~\citep{Tremmel2018,Tremmel2018b}. However, in order for two MBHs to merge, they must evolve down to  separations of $\sim0.001$--$0.01$ pc, where emission of gravitational radiation can bring the MBHs to coalescence. When the MBHs form a binary, on $\sim 1-10$ pc scales, the dynamical evolution is facilitated by complicated interactions with the binary's stellar and gaseous environment~\citep{Quinlan1996,Sesana2015,Vasiliev2015,MacFadyen2008,Cuadra2009,Lodato2009,Roedig2011,Nixon2011,Duffel2019,Munoz2019}, as well as with other MBHs through three/four-body interactions~\citep{Bonetti2018b,2019MNRAS.486.4044B}. The exact evolutionary channel of an MBH binary is  partly determined by the morphology and kinematic structure of their host galaxies \citep{Khan2013,Holley-Bockelmann2015}. Each of these dynamical processes effectively results in a delay  between the merger of two dark matter halos/galaxies and the merger of two MBHs (and the resulting gravitational-wave emission). For the larger-scale evolution, semi-analytic models are informed by results from cosmological N-body and hydrodynamic simulations, while the evolution of MBH binaries is informed by detailed binary evolution simulations. 

Indeed, because of their importance to galaxy formation, MBHs have become an integral part of most large-scale cosmological hydrodynamic simulations, which have been able to reproduce MBH and galaxy scaling relations, as well as implement feedback from MBHs that can successfully regulate and quench star formation in massive galaxies \citep{DiMatteo2005,2012MNRAS.420.2662D,Vogelsberger2014, Schaye2015, Volonteri2016,Tremmel2017, Pontzen2017, Nelson2019, Ricarte2019}. However, the poor resolution of these simulations requires relatively simplistic models for MBH formation, such that the earliest phases of growth and mergers often remain unresolved. Because of the complicated physics involved in fully hydrodynamic simulations, they are also generally limited to relatively small volumes with poor statistics. Most importantly, the dynamical evolution of MBHs is often completely ignored even on resolved $\sim$ kpc scales, though there are important exceptions to this \citep{2012MNRAS.420.2662D,Tremmel2017, Hirschmann2014,Pfister2019}. While some cosmological simulations have been used, with significant post-processing, to account for unresolved MBH binary evolution and make unique predictions relevant for gravitational-wave astrophysics \citep[e.g.][]{Blecha2016, Kelley2017, Katz2019}, they can also be useful tools for improving semi-analytic simulations, which, because of their low computational cost, provide better statistics and the ability to test the effects of different model assumptions.

Recently, cosmological hydrodynamic simulations have seen important improvements to how MBHs are modeled, which has led to new results regarding their growth and dynamical evolution. In the {\sc Romulus} simulations \citep{Tremmel2017}, the dynamical evolution of MBHs is self-consistently followed by using a new, sub-grid model accounting for unresolved dynamical friction \citep{Tremmel2015}. Using these simulations, \citet{Tremmel2018} showed that the formation of MBH pairs with separations below a kpc (the precursors to MBH binaries) often occurs after several Gyr of orbital evolution of MBH pairs at kpc-scale separations, with many MBH pairs failing to ever form a binary within a Hubble time \citep{Tremmel2018b}. Furthermore, the inspiral timescale may also be influenced by (sub-)kpc scale structures, such as galactic bars, spirals, and massive star-forming clumps. These might render the orbital decay significantly stochastic, thus enriching the dynamics of MBHs on kpc-scales~\citep[even leading to the possible ejection of infalling MBHs, see e.g. the recent hydrodynamic simulations of][]{Tamburello2017, Bortolas2020}.

Detailed, high-resolution cosmological simulations of MBH formation at high redshift have also shown that supernova (SN) feedback can stunt early MBH growth in low-mass galaxies \citep{2015MNRAS.452.1502D,Habouzit2017}. In this paper, we combine these new results from large-scale simulations with updated models for MBH binary evolution \citep[e.g.][]{Sesana2014,Antonini2015,Bonetti2018b,2019MNRAS.486.4044B} into the semi-analytic model for MBH and galaxy evolution of \citet{Barausse2012}. The goal of this paper is to understand how the combination of different assumptions for MBH formation, early growth, and dynamical evolution affects the gravitational-wave signal expected for LISA and pulsar-timing arrays.

In \S\ref{sec:sam} we summarize the semi-analytic model and its physical ingredients, as well as the different model variations that we run. Our results are presented in \S\ref{sec:cat}, \S\ref{sec:res}, \S\ref{LF}  and \S\ref{sec:stoch}, including MBH merger rates and predicted mass, redshift, and mass ratio distributions for events detectable by LISA, quasar luminosity functions, as well as predictions for pulsar-timing arrays. In \S\ref{sec:conclusions} we summarize  our results and draw our conclusions.

\section{The semi-analytic model}\label{sec:sam}

We describe the synergistic co-evolution of MBH and their host galaxies using the semi-analytic model of \citet{Barausse2012}, with later updates to specific aspects of it described in \citet{Sesana2014}, \citet{Antonini2015}, \citet{Bonetti2018b} and \citet{2019MNRAS.486.4044B}. In the following, we limit ourselves to a brief review of the model, referring the reader to the aforementioned works for more details,
and highlighting the changes with respect to them. Besides slight changes
in the gas cooling, star formation and AGN feedback prescriptions, which
we describe in this section and against which the results are robust, the improvements to the model on which this paper hinges are instead described in \S\ref{new_model1}--\ref{new_model4}.

The model is built on top of a dark matter merger tree constructed with an  extended Press-Schechter algorithm~\citep{Press1974,Parkinson2008} suitably modified to reproduce the results of N-body simulations~\citep{Parkinson2008}. Baryonic structures are then evolved along this dark matter merger tree using  semi-analytic  prescriptions. These structures include a chemically pristine intergalactic medium, which accretes onto
dark matter halos and gets shock-heated to their virial temperature in high-mass systems at low redshift, or  flows into halos on their dynamical time  along cold filaments~\citep{Dekel2006,Cattaneo2006,Dekel2009}. 
Unlike in  \citet{Barausse2012} and other works based on it,
we allow here for a smooth transition between the shock-heating and cold-filament regimes, by using the results of \citet{correa}, based
on hydrodynamic cosmological simulations from the {\sc EAGLE} project,
and particularly their Eqs. (11)--(15) for the  fraction of hot mode gas accretion. 

Cooling of the intergalactic medium and/or its inflow along cold streams then gives  origin to a cold gas phase (an ``interstellar medium'') that eventually undergoes star formation. In more detail, the model tracks the evolution of gaseous/stellar disks and spheroids, with major galaxy mergers and bar instabilities disrupting disks and turning them into spheroids. The star formation is described in spheroids via the prescriptions of
\citet{Barausse2012}, and in disks via those of \citet{Dutton}, who assume
that star formation takes place in dense molecular clouds.\footnote{Note that these
star-formation prescriptions are slightly different from those used
in previous works based on \citet{Barausse2012} -- such as \citet{Sesana2014}, \citet{Antonini2015}, \citet{Bonetti2018b} and \citet{2019MNRAS.486.4044B} -- but our results are robust against these changes.}
Star formation and SN feedback also drive the chemical evolution of the interstellar medium.

On smaller scales, the model accounts for the formation of nuclear star clusters
\citep[from in-situ star formation and/or from migration of globular clusters to galactic nuclei; see][]{Antonini_Barausse2015,Antonini2015}
and the presence of MBHs.  MBHs form from high-redshift seeds and then grow by accretion of nuclear gas and coalescences with other MBHs brought by galaxy mergers. The mass growth rate  of the reservoir of nuclear gas available within the MBH influence radius for accretion and for in-situ nuclear star cluster formation is assumed to be linearly  correlated with star formation in the spheroid~\citep{Granato2004,Lapi2014,Ricarte2019}, i.e. we assume
$\dot{M}_{\rm nucl}=A\dot{M}_{\rm sf,\,sph}$, with the model's calibration against
local and high-redshift observables~\citep[c.f.][and section \ref{LF}  for a list]{Barausse2012,Sesana2014,Antonini_Barausse2015,Antonini2015}  yielding
$A=5\times 10^{-2}$. Accretion of this nuclear gas onto the MBH is assumed to take place on the viscous timescale evaluated at the influence radius~\citep{Sesana2014}, but
is topped off at $\dot{M}_{\rm bh,\,max}=A_{\rm Edd} \dot{M}_{\rm Edd}$, with $\dot{M}_{\rm Edd}$ the Eddington rate,  and $A_{\rm Edd}=1$ or 2 depending on the seeding model (see below).
The MBH evolution also exerts a feedback on the growth of structures (AGN feedback). While \citet{Barausse2012}, and later work based on it, only accounted for AGN feedback by radio jets, in this paper we also consider the effect of radiative feedback~\citep{2017MNRAS.464.1854B}. We model this effect by injecting  $5\%$ of the AGN 
luminosity into the surrounding gas, and then computing the feedback onto the bulge and diffuse gas distribution via Eqs. (42) and (43) of \citet{Barausse2012} (where we therefore replace $L_{\rm jet,QSO}$
with $L_{\rm jet,QSO}+0.05 L_{\rm bh,QSO}$). 

The most crucial aspects of our model for the prediction of LISA event rates and the amplitude of the pulsar-timing array stochastic backgrounds are the seeding mechanism of MBHs, and the timescales on which MBHs coalesce after their host galaxies merge~\citep{Sesana2007b,Klein2016,2019MNRAS.486.4044B}. 
Several physical models for the mass function of MBH seeds at high redshift have been put forward, see e.g. \citet{Latif:2016qau} for a review. Here, we consider two representative scenarios, namely a light-seed (LS) model in
which seeds are provided by the remnants of population III stars forming in high-redshift, low-metallicity environments~\citep{Madau2001}; and a
heavy-seed (HS) model where the seeds are instead formed by the collapse of proto-galactic disks following the onset of bar instabilities~\citep{Volonteri2008}.

In the LS scenario, we account for the mass losses during stellar collapse by
assuming that seeds have mass $\sim 2/3$ of the mass of the         
initial population III star, which we draw randomly from a log-normal distribution function
centered on $300 M_\odot$, with a standard deviation of 0.2 dex, and a
gap between 140 and 260  $M_\odot$ \citep[to account for the fact that
pair-instability SNae are not believed to leave a black hole remnant; see ][]{Heger2002}.
Moreover, following \citet{Volonteri2003}, in this scenario we only seed the most massive halos, forming from the collapse of the 3.5$\sigma$ peaks of the matter density field at high redshifts $z\gtrsim 15$. To ease the possible discrepancy between LS models and the high-redshift luminosity function~\citep{Madau2014}, we allow for mildly super-Eddington MBH accretion ($A_{\rm Edd}\approx2$) in the LS scenario.

For the HS scenario, we adopt the model by \citet{Volonteri2008}, where protogalactic disks are assumed to become unstable when their Toomre parameter gets lower than a critical threshold $2\lesssim Q_c\lesssim 3$.
The resulting seeds have mass related to the properties of the host halos~\citep[c.f.][for details]{Volonteri2008},
but typically of the order of $\sim 10^5 M_\odot$. Following \citet{Volonteri2008}, also  in this scenario
we only seed halos at $z\gtrsim 15$, but we set $A_{\rm Edd}=1$ (unlike for LSs). Note also that in this paper we fix $Q_c=3$, unlike \citet{2019MNRAS.486.4044B}, where $Q_c=2.5$ was used. A higher value of $Q_c$ increases the number of seeds; therefore in this paper we explore a case where HS form abundantly. This has to be taken into account when comparing this paper's results to previous studies using the same semi-analytic model, and also to other semi-analytic models \citep{Sesana2007b,Ricarte2018b,Dayal2019}, where HSs are normally rare.
Note that  \citet{Klein2016}  assume $Q_c=3$, as in this paper. However, the model for
 the evolution of MBH pairs in  \citet{Klein2016} was simplified with respect
to this work, especially when it comes to triple/quadruple MBH interactions~\citep[c.f.][]{2019MNRAS.486.4044B}, and to the model for the intermediate-scale dynamical ``delays'' between galaxy/halo mergers and MBH coalescences (see below).

When it comes to the delays between galaxy and MBH mergers, \citet{Barausse2012,Sesana2014,Antonini2015,2019MNRAS.486.4044B}
employed a rather sophisticated model for the halo merger times, as well as for the
evolution of MBH binaries on scales below a parsec (where stellar hardening, gas-driven migration, formation of triple/quadruple MBH systems and gravitational-wave emission were all accounted for), but lacked a
description for the dynamics of MBH pairs on scales from several kpc down to a few pc (i.e. the time between the onset of the galaxy interaction and the formation of a bound MBH binary).

In more detail, the model of \citet{Barausse2012,Sesana2014,Antonini2015,2019MNRAS.486.4044B} identifies the ``nodes'' of   
the underlying dark matter
merger tree with the epoch at which
the smaller (``satellite'') halo first enters the host halo.
The satellite halo is then assumed to survive within the host as a subhalo,
while losing mass because of tidal stresses/evaporation and sinking toward the center under the effect of dynamical friction.
We model the mass loss due to tidal effects by following \citet{Taffoni2003},  while for the dynamical-friction time we
adopt the fit (to N-body simulations) of \citet{Boylan-Kolchin2008}. Only once the dynamical friction time has elapsed, is the subhalo 
assumed to lose its individual identity. At that point, \citet{Barausse2012,Sesana2014,Antonini2015,2019MNRAS.486.4044B} assume that the satellite and host galaxies merge,
and that  their MBHs, if present, form a binary.

 The evolution 
of MBH binaries was then followed by accounting (via semi-analytic prescriptions) for the three-body interactions with stars~\citep{Quinlan1996,Sesana2015}, which extract energy and angular
momentum from the binary. This results in a slow secular ``hardening'' on timescales $\sim$ Gyr
from separations comparable to the hardening radius  $a_h = G m_2/(4\sigma^2)$ \citep[where $m_2$ is  the secondary MBH's mass, see][]{Quinlan1996} down to the separation
(typically $10^{-2}-10^{-3}$ pc) at which gravitational-wave emission is sufficient to drive the MBHs to merger in less than a Hubble time. 
In gas-rich galactic nuclei,
where the nuclear gas mass contained within the influence radius of the binary exceeds the total binary mass, the 
model assumes instead that the binary's evolution is driven by the gas on its viscous timescale $\sim 10^7-10^8 $ yr, again down to the separation at which  gravitational-wave emission becomes dominant.
Finally, first in \citet{Antonini2015,Klein2016} and then, in a more rigorous fashion in \citet{Bonetti2018b,2019MNRAS.486.4044B}, the model accounts  for the possible formation of triple/quadruple MBH systems, which can form when two galaxies merge before the hosted MBH binary (or binaries) has had time to coalesce. These MBH triple/quadruple systems can trigger the merger  of at least two MBHs via hierarchical Kozai-Lidov interactions~\citep{Kozai1962,Lidov1962},   or chaotic
three-body interactions~\citep{Bonetti2018a}.

\subsection{Delays due to the dynamical evolution of merging galaxies}\label{new_model1}

The timescale of \citet{Boylan-Kolchin2008}, which as mentioned above we use to model the
dynamical friction from the satellite subhalo within its host, correctly accounts for
the evolution of the dark matter in a galaxy merger, but may be too short to account also for the subsequent  dynamics of the baryonic components (and particularly the MBHs). We therefore augment it by several additive terms, accounting for the dynamical friction between galaxies, and later for the dynamical friction exerted on the individual MBHs (naked or still surrounded by a core of stars)  by
the  stellar background, down to the hardening radius.

In more detail, on the scale of interacting galaxy pairs, we describe the dynamical friction timescale of two galaxies by following \citet{Binney2008}:
\begin{eqnarray}\label{galdf}
T_{\rm df}&=&\frac{2.7\, {\rm Gyr}}{\ln \Lambda} \frac{R_i}{30 \,{\rm kpc}}\left(\frac{\sigma_{\rm h}}{200\, {\rm km/s} }\right)^2\left(\frac{100\, {\rm km/s}}{\sigma_{\rm s}}\right)^3 \\
\Lambda^{-1}&=&\max\left[\frac{\sigma_{\rm s}}{2^{3/2}\sigma_{\rm h}},\sqrt{2}\left(\frac{\sigma_{\rm s}}{\sigma_{\rm h}}\right)^3\right]\label{galdf2}
\end{eqnarray}
where $\sigma_{\rm s}$ and $\sigma_{\rm h}$ are the velocity dispersions of the satellite and host galaxies, while $R_i$ is the initial separation, which we set to the  half-light radius of the galaxy that will form from the merger~\citep[c.f. Sec.~2.2.2 of][]{Barausse2012}.
Note that this expression accounts for the progressive tidal stripping of the secondary galaxy along the evolution~\citep[][at least as long as both galaxies are modeled
simplistically as isothermal spheres]{Binney2008}. 
However, when considering the earlier dynamical-friction-driven 
evolution of the satellite subhalo within its host, if the satellite galaxy has already been
completely 
disrupted by tidal stripping and evaporation
\citep[which we describe by the model of][]{Taffoni2003},
then applying Eqs.~\eqref{galdf}--\eqref{galdf2} is not physically meaningful.
For those systems, we therefore replace  Eqs.~\eqref{galdf}--\eqref{galdf2}
by the dynamical friction timescale of a ``naked'' MBH~\citep{Binney2008}:

\begin{equation}\label{galdf3}
T_{\rm df}=\frac{19\, {\rm Gyr}}{\ln[R_i\sigma_{\rm h}^2/(G M_{\rm bh,sat})]}\left(\frac{R_i}{5 \,{\rm kpc}}\right)^2 \left(\frac{\sigma_{\rm h}}{200\, {\rm km/s} }\right) \left(\frac{10^8  M_\odot}{M_{\rm bh,sat}}\right)
\end{equation}
where again $R_i$  is set to to the  half-light radius of the host galaxy (because the satellite galaxy has been completely destroyed).

\subsection{Delays due to dynamical evolution of MBH pairs on kpc scales}\label{new_model2}

To account for the possibility that simplified prescriptions such as Eqs.~\eqref{galdf}--\eqref{galdf3} may not be sufficiently realistic on smaller scales, we also define an additional timescale $T_{\rm Romulus}$ designed to
fit the times spent by MBH binaries at separations at hundreds of pc (or larger) in the {\sc Romulus} simulation of \citet{Tremmel2017}.
In more detail, to compute $T_{\rm Romulus}$ for each galaxy merger, we sample a probability distribution function
\begin{multline}
      \frac{{\rm d}p}{{\rm d}\log T_{\rm Romulus}}=\frac{F}{\sqrt{2\pi} \sigma} e^{-\frac{(\log T_{\rm Romulus}-\log\mu)^2}{2 \sigma^2}}\\+(1-F) \delta(\log T_{\rm Romulus}-\log t_{0})\,, \label{romulus_pdf}
\end{multline}
which is the linear combination of a log-normal distribution centered on $\mu$ and with standard deviation $\sigma$, and a Dirac delta peaked at the age of the universe $t_{0}=13$ Gyr.
This bimodal distribution models the fact that only a fraction $F<1$
of galaxy mergers results in a close MBH pair in the simulation 
of \citet{Tremmel2018}, as shown in their Fig.~1 (from which we extract $F$ as function of the stellar mass of the primary galaxy).
The times spent by \textit{close} MBH pairs at separations of hundreds of pc
are instead shown in Fig.~6 of \citet{Tremmel2018} as the differences between the orange and blue points.
We fit these differences with a log-normal distribution, which yields $\mu=1.3$ Gyr and  $\sigma=0.7$ (dex).

In order to avoid double counting the galaxy merger timescale naturally present in the cosmological simulations, we assume that the time spent by MBH binaries at separations of 100s-1000s pc to be $\max(T_{\rm df},T_{\rm Romulus})$, where T$_{\rm df}$ is the galaxy dynamical friction timescale discussed in the previous section.\footnote{The reader may think that because the {\sc Romulus} simulation self-consistently describes the dynamics of MBH pairs, one may  simply  use $T_{\rm Romulus}$ in  place of $T_{\rm df}$. However, this would not be a robust prescription,
as our modeling of the  {\sc Romulus} results  via Eq.~\eqref{romulus_pdf} is purely stochastic. For instance, an MBH whose host galaxy has been tidally disrupted and for which a
long $T_{\rm df}$, given by Eq.~\eqref{galdf3}, would be in order, may be assigned a short $T_{\rm Romulus}$ by sampling Eq.~\eqref{romulus_pdf}. Therefore, to remain conservative on our merger rates, we use the maximum of these values.}

\subsection{Dynamical friction on bound MBH pairs (below 100 pc scales)}\label{new_model3}

At separations comparable with the influence radius of the primary MBH, the MBHs become bound in a binary, which changes the details of the dynamical friction~\citep{Dosopoulou2017}.
The dynamical friction timescale for the binary's evolution from the primary's influence radius $r_{\rm infl}$ down to a smaller separation $\chi r_{\rm infl}$ is given by~\citep{Dosopoulou2017}
\begin{align}\label{bhdf}
    T^{\rm df,infl}_{\rm bare}&=1.5\times 10^7
    \dfrac{\left[\ln \Lambda' \alpha + \beta+\delta\right]^{-1}}{(3/2-\gamma)
    (3-\gamma)} \left(\chi^{\gamma-3/2}-1\right)\nonumber\\
    &\left(\dfrac{M_1}{3\times 10^9 M_\odot}\right)^{1/2}
    \left(\dfrac{M_2}{10^8 M_\odot}\right)^{-1}
    \left(\dfrac{r_{\rm infl}}{300\,\rm pc}\right)^{3/2} \rm yr ,
\end{align}
where $M_1$ and $M_2$ are the primary and secondary MBH masses, and we assume for simplicity $\Lambda'\approx r_{\rm infl}\sigma^2/(G M_2)\approx M_1/M_2$.
The coefficients $\alpha$, $\beta$ and $\delta$ are functions of the power-law exponent  $\gamma$ of the stellar density near $r_{\rm infl}$, i.e. $\rho_{\star}\propto (r/r_{\rm infl})^{-\gamma}$, and are given by Eqs.~21--23 of~\citet{Dosopoulou2017}. Here, we assume $\gamma=1$. 

Because Eq.~\eqref{bhdf} does not account for the fact that some stellar mass from the satellite galaxy (if it has not been completely disrupted by tidal effects earlier on) can remain bound to the secondary MBH even within $r_{\rm infl}$, we follow again \citet{Dosopoulou2017} and model this effect by the timescale
 \begin{align}\label{eq:bhdf_dress}
  T^{\rm df,infl}_{\rm dressed}&=1.2\times 10^7 \dfrac{\left[\ln \Lambda \alpha + \beta+\delta\right]^{-1} 
 }{(3-\gamma)^2} \left(\chi^{\gamma-3}-1\right)\nonumber\\
 &\left(\frac{M_1}{3\times 10^9 M_\odot}\right)
  \left(\dfrac{100\,\rm km\ s^{-1}}{\sigma_s} \right)^3\  
 \rm yr \,,
\end{align}
where $\Lambda$ is given by Eq.~\eqref{galdf2}. 
We therefore assume that the decay timescale from  the influence radius is
given by
\begin{equation}\label{tdf-rinf}
 T^{\rm df,infl}=\min\left(T^{\rm df,infl}_{\rm bare},T^{\rm df,infl}_{\rm dressed}\right)\ .
\end{equation}

 \begin{table*}
    \centering
     \caption{Summary of all models explored in this work.}
     \begin{tabular}{|l||c|c|c|c|}
     \hline 
    \multirow{2}{*}{Model}  &  SN feedback on MBH growth   & galaxy/BH dynamical friction  &  kpc-scale delays & $\lesssim 100$ pc delays   \\
     &   (Habouzit et al. 2017)  &  (Binney \& Tremaine 2008) & (Tremmel et al. 2018) &  (Dosopoulou \&  Antonini 2017)  \\
    \hline \hline
  
    SN-delays & \cmark & \cmark & \cmark & \cmark  \\\hline
    SN-delays-medium & \cmark & \cmark & \xmark & \cmark  \\\hline
    SN-delays-short & \cmark & \xmark & \xmark & \cmark  \\\hline
    SN-nodelays & \cmark & \xmark & \xmark & \xmark  \\\hline \hline
    noSN-delays & \xmark & \cmark & \cmark & \cmark  \\\hline
    noSN-nodelays & \xmark & \xmark & \xmark & \xmark  \\
    \hline
  \end{tabular}
  \label{tab:models_new_names}
 \end{table*}

\subsection{Effect of SN feedback on MBH growth}\label{new_model4}    

Another ingredient that we add to our semi-analytic model is the possibility that
SN feedback may stunt the growth of MBH in low-mass galaxies. Indeed, while the effect of SN feedback on star formation is already included in the model of \citet{Barausse2012},  its effect on the MBH growth had not been included yet. In fact, for MBH growth the relevant quantity is not the overall gas fraction in the galaxy, but the physical state (density, temperature) of the gas near the MBH and how this gas is distributed \citep{2015MNRAS.452.1502D}.
 
The exact way SN explosions affect gas, and therefore whether they are able to evacuate the gas near MBHs, strongly depends on the details of the process, i.e. on how the energy released in the explosion couples to the gas. For instance, \citet{Habouzit2017} find that weak thermal and kinetic SN feedback do not have a dramatic effect on the MBH growth, while in SN feedback models where gas cooling is delayed because of the (unresolved) shocks,  accretion onto MBHs is suppressed in low-mass systems. Importantly, the observational properties of galaxies and MBHs are well reproduced in the simulations of \citet{Habouzit2017} only with the latter implementation of SN feedback.
To account for the possible suppression of MBH growth in low-mass galaxies caused by SNae, we assume that 
the growth rate $\dot{M}_{\rm nucl}$ of the nuclear gas reservoir from which the MBH accretes is quenched in galaxies where the escape velocity from the spheroidal bulge is lower than 270 km/s. The latter is indeed the speed of SN winds in the delayed cooling simulations of \citet{Habouzit2017}. This suppresses MBH growth and lengthens the viscous timescale for the evolution of MBH binaries in circumbinary disks.

\section{Catalogs of merging MBH binaries}\label{sec:cat}

Based on the output of our semi-analytic model, we compute the
expected detection rate of merging MBH binaries by the LISA mission.
In more detail, we consider 
the models  in Table~\ref{tab:models_new_names}, and for each of them we produce synthetic catalogs of merging MBH binaries, including all relevant information on each binary, such as the component masses, merger redshift and spins, 
as well as the properties of the host merging galaxies (e.g. their merger redshift, their masses, etc.). 
The merger rate per unit redshift is calculated by summing the contributions within each redshift bin $\Delta z$:
\begin{equation}
 \frac{d^2N}{dz dt} = \frac{4\pi c}{\Delta z}\sum_{N\in \Delta z}W(z)\left[\frac{d_L(z)}{1+z} \right]^2
\end{equation}
where $d_L$ is the luminosity distance, and $W(z)$ is the comoving number density of the binary (obtained from the comoving number density of its host galaxy).

For each MBH binary we then calculate the signal-to-noise ratio (SNR) $\rho$, averaged over polarization, inclination and sky position, i.e~\citep[see e.g.][]{Cornish2018}:
\begin{equation}
 \rho^2 = 4\int_{f_{\rm min}}^{f_{\rm max}} \frac{|h(f)|^2}{S_n(f)}df\,,
\end{equation}
where $h(f)$ is the gravitational-wave strain amplitude in Fourier space, and $S_n(f)$ is the LISA sensitivity. We calculate the gravitational-wave strain using the \emph{PhenomC} inspiral-merger-ringdown model \citep{Santamaria2010}, and we use the  LISA sensitivity curve from \citet{LISA2017}, including the contribution from the foreground from Galactic binaries~\citep{2017JPhCS.840a2024C,Cornish2018}. The time to merger is drawn from a uniform distribution between 0 and the nominal mission duration (4 yr). We use $\rho=8$ as the detection threshold. The number of sources detectable in a 4-yr LISA mission  is shown in Table~\ref{tab:rates_new_names}, and will be discussed below.

\section{LISA detection rates: delay mechanisms and SN feedback}\label{sec:res}

In this section, we discuss in detail the results of our various models for the predicted merger rates of MBHs and their potential detection with LISA. Our models are summarized in Table \ref{tab:models_new_names}. Each model incorporates a different combination of prescriptions pertaining to the intermediate-scale dynamical evolution of MBHs (as described in \S\ref{new_model1}--\ref{new_model3}), as well as SN-regulated MBH growth (\S\ref{new_model4}). 
Each evolutionary model is applied starting from either an LS or HS high-redshift population, as described in \S\ref{sec:sam}.
 
In Table \ref{tab:rates_new_names}, we report the total number of MBH mergers, as well as the number of detections expected in $4$ yr of observations with LISA, for each model. For the noSN-delays model, we find the same general trend found in previous work \citep{Sesana2007b,Klein2016,2019MNRAS.486.4044B}, i.e. that common, low-mass seeds produce significantly more detected mergers than high-mass seeds. Interestingly, this trend is reversed for models accounting for the effect of SN feedback on MBH growth. 
Low mass black holes are those whose merger and detection prospects are most affected by SN feedback, because their arrested growth yields not only longer binary formation timescales (see Eqs.~\ref{galdf3} and \ref{bhdf}), but also lower SNRs, as low-mass binaries tend to coalesce at the high-frequency end of the LISA sensitivity curve.

The impact of SN explosions on the HS models is more subtle. Without any intermediate-scale delays, SN feedback has little effect, as the MBHs are seeded at masses large enough to form binaries in a timely manner and in a mass range readily detectable with LISA. However, when delay times to MBH binary formation are included, SN feedback increases the merger rate. Without SN regulation, massive seeds experience more early growth,   
generally increasing their spins~\citep{Barausse2012,Sesana2014}. Merging MBHs therefore experience stronger recoil kicks, which tend to remove 
the merger remnant from the center of the host galaxy. This obviously  prevents the remnant from coalescing with other MBHs brought in by future galaxy mergers. Without the delay mechanisms in place for MBH binary formation, MBHs merge before they have a chance to grow and substantially spin up, irrespective of whether SN feedback is present or not.

High-mass seed models are much more affected by the inclusion of MBH dynamical evolution. The inclusion of even one of the delay timescales discussed in \S\ref{sec:sam} results in an order of magnitude decrease in both the total and detected MBH merger rate. More detailed choices of additional delay mechanisms have a reduced effect, but can still be important. In particular, the inclusion of kpc-scale dynamical evolution of MBHs within galaxy merger remnants, as predicted from cosmological simulations \citep{Tremmel2018}, results in a factor of $\sim3$ decrease in MBH mergers for HS models. LS models are much less sensitive to these intermediate-scale delays, but also see a decrease of a factor of $\sim2$ when including these galaxy-scale delays.

 \begin{table}[h!]
  \centering
  \caption{Total number of sources and detections expected in $4$ yr of observation with LISA for all the models explored here (see Table \ref{tab:models_new_names} for a summary).}
  \label{tab:landscape}
  \begin{tabular}{|l||c|c|c|c||c|c|}
    \hline 
    \multirow{2}{*}{\textbf{Model}}  &  \multicolumn{2}{c|}{\textbf{LS}} & \multicolumn{2}{c||}{\textbf{HS}}   \\
    \cline{2-5}
                    & \textbf{Total} & \textbf{Detected} & \textbf{Total} & \textbf{Detected} \\
    \hline \hline
    \multicolumn{3}{c}{\emph{SN feedback}} \\ \hline
    SN-delays & 48 & 16 & 25 & 25 \\\hline
    SN-delays-medium & 157 & 38 & 89 & 88  \\\hline
    SN-delays-short & 169 & 33 & 74 & 73  \\\hline
    SN-nodelays & 178 & 36 & 1269 & 1269  \\\hline\hline
      
    \multicolumn{3}{c}{\emph{No SN feedback}} \\ \hline
    noSN-delays & 192 & 146 & 10 & 10 \\ \hline
    noSN-nodelays & 1159 & 307 & 1288 & 1288   \\ \hline

    \hline
  \end{tabular}
  \label{tab:rates_new_names}
 \end{table}

In the following sections, we describe in more detail the effect of SN feedback and intermediate-scale dynamical delay timescales on the predicted distribution of MBH merger properties and their astrophysical implications. We note that  even in our most pessimistic models, we predict several detectable mergers within the LISA nominal mission duration of 4 yr, similar to previous work~\citep{Klein2016,Dayal2019,2019MNRAS.486.4044B}. Our prediction is higher than those derived from the Illustris cosmological simulation \citep{Katz2019}, though this is likely due to their limited resolution and lack of MBHs in low-mass dwarf galaxies \citep{Volonteri2020}.

\begin{figure}
\centering
\includegraphics[width=0.45\textwidth]{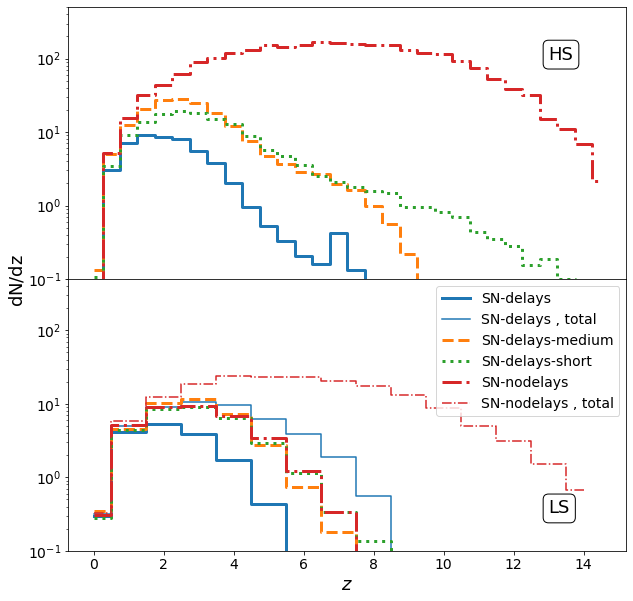}
\caption{Number of total (thin lines) and detected (thick lines) MBH mergers per unit redshift during a 4-year LISA mission in the HS and LS models that include SN feedback (upper and lower panels, respectively; see Table \ref{tab:models_new_names} for a summary of the models). The difference between total and detected mergers for HS models is very small and not visible in this plot.}
\label{fig:SN_dNdz}
\end{figure}

\subsection{Delay mechanisms}\label{5.1}

We first focus on the effect of MBH binary formation delay timescales. Beyond standard prescriptions for dark matter halo sinking timescales,  for MBH binary hardening/gas-driven migration, and for triple/quadruple MBH interactions,  which our model already includes ~\citep{Barausse2012,Sesana2014,Antonini2015,2019MNRAS.486.4044B}, we incorporate three different  intermediate-scale delay mechanisms in this work (c.f. \S\ref{sec:sam}). These are to be added to the aforementioned delays already included in the model. As outlined in Table \ref{tab:models_new_names}, the `delays-short' models account for the dynamical friction on bound MBH pairs beginning at the primary MBH's sphere of influence \citep{Dosopoulou2017}. The `delays-medium' models additionally account for dynamical friction acting on merging galaxies \citep{Binney2008}. Finally, the `delays' models further incorporate a third delay timescale associated with the dynamical evolution of MBHs within galaxies, which has been calibrated to cosmological simulations \citep{Tremmel2018}.
In this section, all of the models include SN-regulated MBH growth.  Note also that the `nodelays' model, which we include as baseline, may be physically relevant to describe a scenario where stellar hardening is efficient
at driving the evolution of MBH binaries already at separations of hundred of pc~\citep{Sesana2006,Khan2012b,Khan2016}.

Figure~\ref{fig:SN_dNdz} shows the redshift distribution of the MBH mergers detectable during LISA's 4 yr of nominal mission duration. The top panel plots the results for HS models. Without intermediate-scale delays, there are a large number of MBH mergers predicted, the majority of which occur at high redshift ($z>5$). However, the inclusion of even one of these intermediate-scale delays (i.e. the SN-delays-short model; green line) shows a substantial decrease in mergers at $z>5$. The addition of galaxy merger dynamical friction timescales (SN-delays-medium; orange line) has a small effect of shifting high-$z$ MBH mergers to lower redshift, though the difference between SN-delays-medium and SN-delays-short is much less significant that that between SN-nodelays and SN-delays-short. A more significant difference is seen when the delays from MBH dynamical evolution on kpc scales is included (SN-delays; blue line), which results in an additional decrease in the merger rates, particularly at $z>2$.

For low-mass seeds (bottom panel of Fig.~\ref{fig:SN_dNdz}), only when kpc-scale delays are included (SN-delays; blue line) is there an appreciable effect on the MBH merger rate. Regardless of the considered binary formation delay timescale, many of the MBH mergers from low-mass seeds will be missed by LISA (as can be seen by comparing to the {\it total} number of MBH mergers, shown by the dashed lines for two of the LS models). This is because, unlike the HS models, LSs result in many MBH mergers with total mass low enough to give SNRs below our detection threshold 
($\rho=8$).

\begin{figure}
\centering
\includegraphics[width=0.45\textwidth]{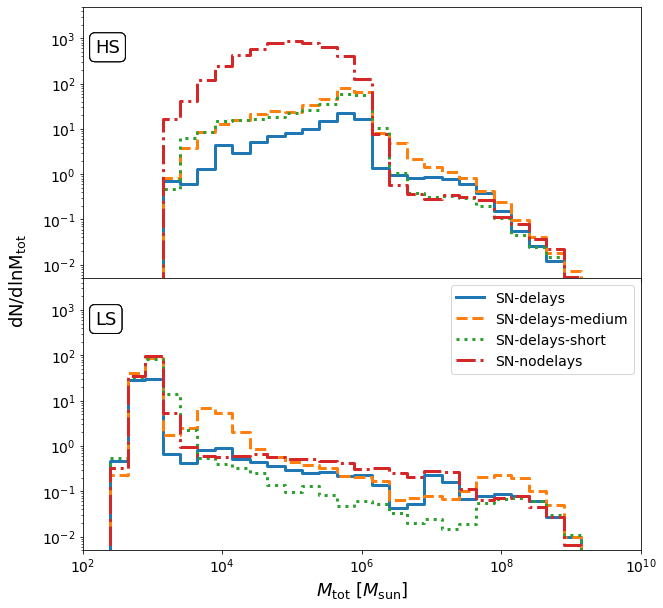}
\caption{Number of detected mergers as a function of MBH-binary mass in the HS and LS models that include SN feedback (upper and lower panels, respectively). The mass distribution of LS models is barely unaffected by time delays. In HS models, low-mass (and high-redshift; see Fig.~\ref{fig:SN_dNdz}) binaries acquire large delays due to dynamical friction. 
}
\label{fig:SN_dNdM}
\end{figure}

Figure~\ref{fig:SN_dNdM} shows the total mass distribution of detected MBH mergers for both HS (top) and LS (bottom) models. The inclusion of intermediate-scale delays mostly affects the number of low-mass mergers, which typically take place as the result of low-mass, high-$z$ galaxy mergers. For HSs, this means that longer delay timescales result in fewer mergers of mass $10^4-10^6 M_{\odot}$. For LSs the effect is less pronounced, but the decrease in the overall merger rate seen in Table \ref{tab:landscape} and Fig.~\ref{fig:SN_dNdz} is also due to an overall decrease of mergers with mass below $10^4 M_{\odot}$. 

\begin{figure}
\centering
\includegraphics[width=0.45\textwidth]{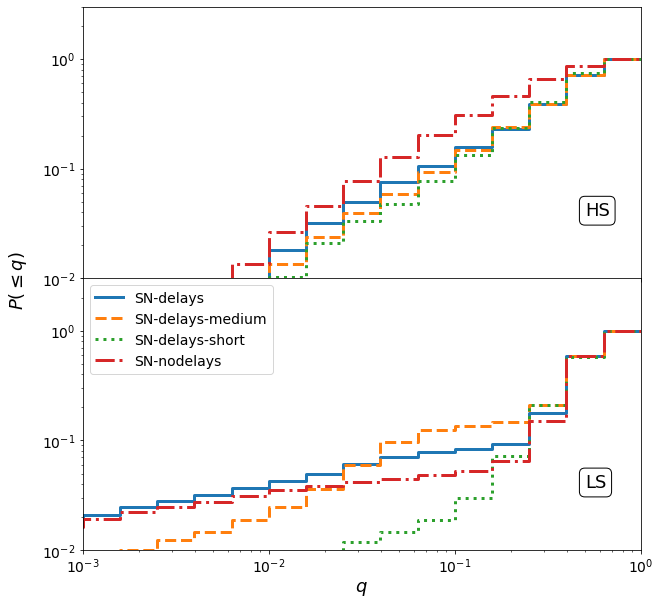}
\caption{Fraction of detected mergers with mass ratio below $q$ in the HS and LS models with SN feedback (upper and lower panels, respectively). The distribution in HS models slightly shifts toward larger $q$ (more equal mass ratio) in models with large delays (note that about $10\%$ of the mergers in the HS models have $q\lesssim 0.1$). In contrast, including all time delays in LS models produces a distribution with a pronounced tail toward low $q$.}
\label{fig:SN_models_dNdq}
\end{figure}

Figure~\ref{fig:SN_models_dNdq} shows the cumulative distribution of MBH merger mass ratios $q=M_2/M_1\leq1$
for events detectable with LISA. The distribution is largely unaffected by the MBH binary delay timescale for HSs, except for a slight steepening trend toward higher mass ratios when delays are included. The merger mass ratio for the LS models is much more affected by binary formation delays, but interestingly our model incorporating the longest binary formation timescales (SN-delay) is more similar to the model without delays (SN-nodelays) than to the short and medium delay models. This is caused by the implementation of dynamical friction on bound MBH pairs, which has an explicit dependence on $M_2^{-1}$ (see Eq.~\eqref{bhdf}): this will preferentially delay low-mass-ratio MBH mergers, which have already longer sinking timescales \citep[see, e.g.,][]{2020MNRAS.493.3676O}, resulting in fewer mergers with low mass ratios. 
Dynamical friction between galaxies delays  galaxy mergers and allows the MBHs (and particularly the primary) to grow further prior to binary formation, resulting in more low-mass-ratio mergers. The additional delay associated with kpc-scale dynamical evolution increases this effect until the distribution is nearly the same as that without any delays.

\subsection{SN feedback}

\begin{figure}
\centering
\includegraphics[width=0.45\textwidth]{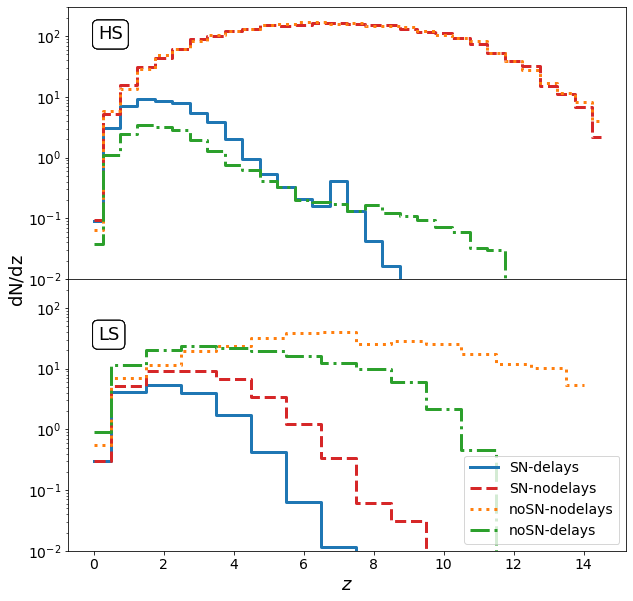}
\caption{Number of detected mergers per unit redshift during a 4-year LISA mission in the HS and LS models (upper and lower panels, respectively; see Table \ref{tab:models_new_names} for a summary of the models). Note that the HS models with no delays (both with and without SN feedback) are almost indistinguishable.}
\label{fig:SNyn_dNdz}
\end{figure}

Next, we explore the effect of SN feedback~\citep{Habouzit2017} on the detection rates and properties of MBH binaries. We compare this effect to that of binary formation delay times by including both `delays' and `nodelays' models, each with and without SN feedback. In Fig.~\ref{fig:SNyn_dNdz}, we show the redshift distribution of the predicted 4-year LISA detections. As also evident from Table~\ref{tab:rates_new_names}, the HS model is not strongly affected by SN feedback, while the low-mass seeds are significantly impacted (compare the pink and red curves in Fig.~\ref{fig:SNyn_dNdz}).

\begin{figure}
\centering
\includegraphics[width=0.45\textwidth]{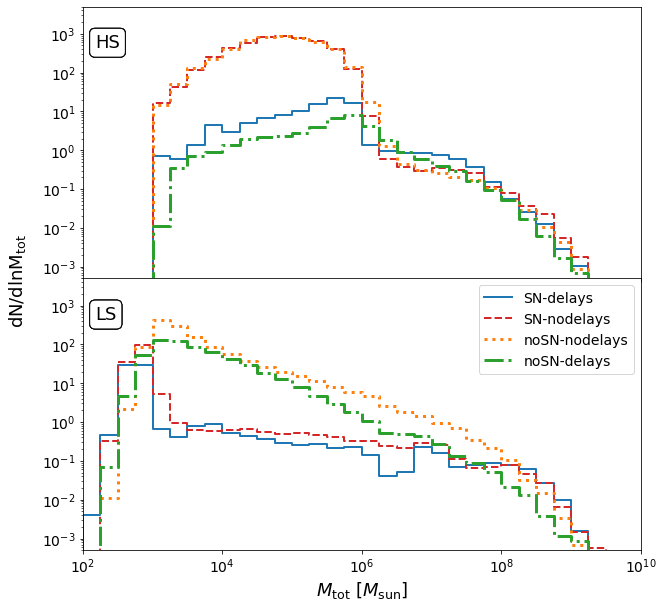}
\caption{Number of detected mergers as a function of total MBH mass in the HS and LS models (upper and lower panels, respectively). The mass distribution of LS models is practically unaffected by time delays. In HS models, low-mass (and high-redshift; see Fig.~\ref{fig:SN_dNdz}) binaries are ejected from their host halos in models that include time delays.}
\label{fig:SNyn_dNdM}
\end{figure}

The reason for this sensitivity of LS models to SN feedback can be understood from Fig.~\ref{fig:SNyn_dNdM}, which shows that the number of detected mergers in the $10^5-10^7M_{\odot}$ mass range, to which LISA is most sensitive, is drastically reduced relative to models without SN feedback. This deficit is  inherited from the fact that the \emph{total} number of these binaries (i.e. before the SNR cutoff) is reduced in LS models with SN feedback,
because SN winds tend to expel nuclear gas,
which results in suppression of  seed growth (due to lack of accretion) and in longer timescales for
binary migration in circumbinary disks.

\begin{figure}
\centering
\includegraphics[width=0.45\textwidth]{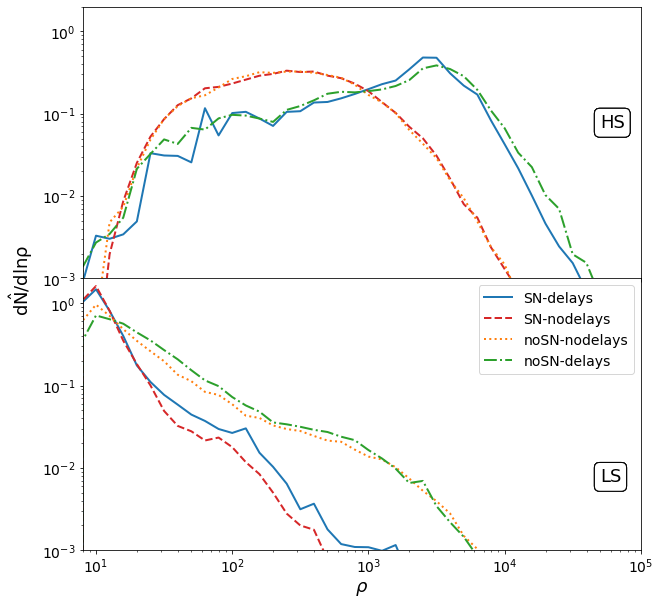}
\caption{Distribution  (normalized to the total number of detected events) of SNR in the HS and LS models (upper and lower panels, respectively). The distribution of SNR in the LS models is very mildly affected by dynamical effects that cause time delays, whereas in the HS case, dynamical effects on $100$-pc scales affect mostly low-mass systems (see Fig.~\ref{fig:SN_dNdM}) and/or systems with unequal mass ratios (see Fig.~\ref{fig:SN_models_dNdq}). As a result, the 'surviving' binaries have higher SNR in HS models with large time delays. Note also that the SNR in LS models without SN feedback shift toward slightly higher values relative to the case with SN feedback.}
\label{fig:SN_dNdlSNR}
\end{figure}

Next, in Fig.~\ref{fig:SN_dNdlSNR}, we compare the shape of the distribution (normalized to total detections) of SNR values  when the four antipodal model prescriptions, delay/no-delay and SN/no-SN, are considered. As in the previous figures, the upper panel shows results from the HS scenario, while the bottom panel concerns the LS case. From the figure, it can be seen that the SNR distribution reflects the mass distribution of the sources (Fig.~\ref{fig:SNyn_dNdM}). As expected, the SNR distribution in LS models peaks at lower SNR values due to the low source mass. The shape of the mass distribution for LSs is not greatly affected by dynamical delays (see \S\ref{5.1}), but is significantly affected by SN feedback. As a result, when SN explosions are included, the number of high-SNR sources in  the LS scenario declines. For HS models, the SNR distribution is more affected by MBH dynamics, with longer MBH binary formation timescales resulting in more 
numerous
high-SNR systems  than low-SNR ones. This is due to the decrease in low-mass MBH mergers, to the delays allowing for the  MBHs (and especially the primary) to grow prior to binary formation and merger, and to the mergers being delayed to lower redshift, where their SNR is naturally higher (see Fig.~\ref{fig:SN_dNdz}). Note that the more massive MBHs in the HS model are much less affected by SN winds because of their high mass, which is already in the range to which LISA is most sensitive.

Given the steeply growing distribution toward lower SNRs in the LS models, an important conclusion from Fig.~\ref{fig:SN_dNdlSNR} is that, if LSs are indeed realized in the universe, the total number of LISA detections may be extremely sensitive to the SNR threshold or, equivalently, to the LISA noise budget  \citep[see also][]{2019MNRAS.486.4044B}.
Therefore, in view of the small number of expected detections per year in the LS scenario (in which the inclusion of SN feedback is crucial), particular care should be put into optimizing LISA's sensitivity at high frequencies, as any degradation can translate into significant event losses.
For instance, changing the single link optical measurement system noise from
$10 {\rm pm}/\sqrt{\rm Hz}$, adopted in \citet{LISA2017}, 
to $15 {\rm pm}/\sqrt{\rm Hz}$, in order to account for the margin on this noise contribution inserted in 
the LISA Science Requirements Document~\citep{SRD},
does not affect the detection rates for the HS models, but LS detection rates are reduced by a factor of $\sim2$. To improve the sensitivity to low-mass systems, joint observation campaigns with space-based missions that are scheduled at the same time as LISA -- e.g. TianQin~\citep{tianqin} -- and which are also sensitive to MBH binaries~\citep{tianqinMBH,tianqinRD} may be particularly useful.

\subsection{Astrophysical Implications}

In this section we examine our model's predictions
for the MBH binary population in terms of the properties of their host galaxies and their galaxy mergers. These predictions are crucial for interpreting LISA detections in the context of galaxy formation and evolution, so as  to use LISA data to constrain models for the synergic co-evolution of MBHs with galaxies~\citep{Sesana2007b, Volonteri2009, Berti2008, Klein2016, Ricarte2018b, 2019MNRAS.486.4044B}, and to attempt multi-wavelength follow-up observations of LISA sources~\citep{Tamanini2016}.

\subsubsection{Host galaxies of MBH mergers}

\begin{figure}
\centering
\includegraphics[width=0.45\textwidth]{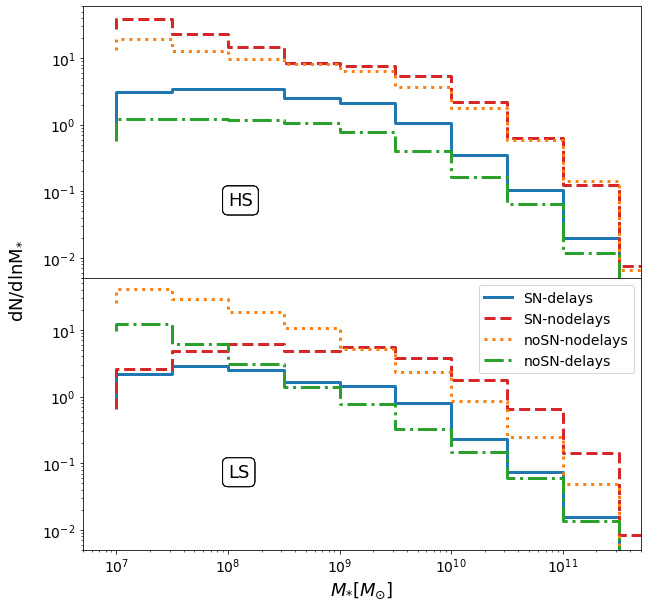}
\caption{Number of detected MBH mergers per unit stellar mass of the host galaxy in the HS and LS models (upper and lower panels, respectively). MBH merger rates are suppressed in low-mass halos by SN winds, and are everywhere suppressed by delays due to MBH dynamics on 10s-1000s pc scales. Regardless of the model's details, a consistent prediction is that the majority of MBH mergers detectable by LISA occur in dwarf galaxies of mass M$_{\star}\lesssim10^{9.5} M_{\odot}$.
}
\label{fig:SNyn_dNdMhalo}
\end{figure}

In Fig.~\ref{fig:SNyn_dNdMhalo}, we examine how both delays and SN feedback affect the distribution of the host galaxy stellar mass. Once again, HS models are not strongly affected by SN feedback, but are sensitive to the dynamics of MBH binary formation. The inclusion of delays to MBH formation decreases the number of detected mergers predicted by the model, without greatly affecting the shape of the host galaxy distribution. Conversely, the LS models are affected by both delays and SN feedback. The inclusion of delays (in particular the kpc-scale dynamical evolution of MBHs pairs) decreases the number of MBH mergers in higher mass galaxies, while SN feedback decreases the number of mergers at low masses (M$_{\star} \lesssim 10^8 M_{\odot}$). The result is that models that include SN feedback have flatter galaxy mass distributions at the low-mass end. This is due to the fact that SN winds are efficient at removing gas  from the shallow potential wells of low-mass galaxies, thus effectively shutting off accretion on LSs -- which cannot reach the masses at which LISA is most sensitive ($\sim 10^5$--$10^7M_\odot$) --,
and resulting also in longer binary migration timescales 
in circumbinary disks.

It is important to note that all models, despite their large differences in merger rates and mass distributions, predict that mergers in low-mass galaxies (M$_{\star} \lesssim 10^{9.5} M_{\odot}$) dominate the MBH merger population that will be detected by LISA. This is in agreement with previous cosmological simulations~\citep{Volonteri2020}. It is therefore critical that models that hope to make predictions relevant for LISA fully resolve high-redshift mergers between low-mass galaxies, a particularly difficult challenge for large-scale cosmological simulations.

\subsubsection{The (dis)connection between galaxy and MBH mergers}

\begin{figure}
\centering
\includegraphics[width=0.45\textwidth]{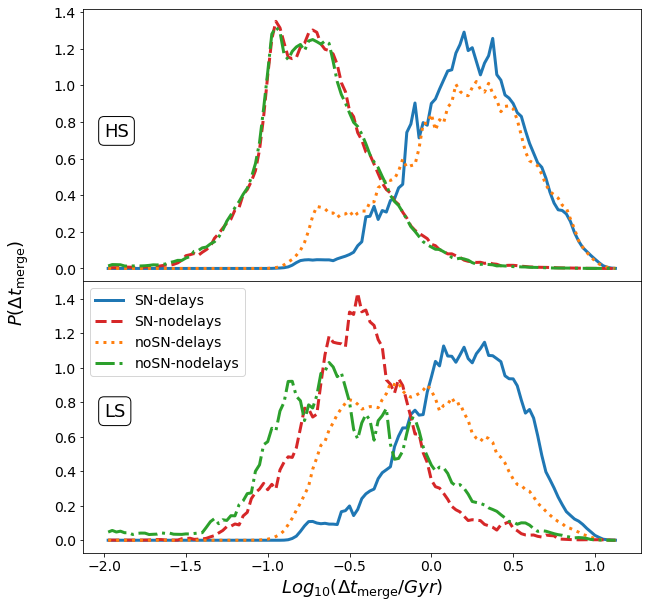}
\caption{Distribution of delay times between galaxy and binary merger for the detected binaries, in the LS and HS models.}
\label{fig:DeltaT}
\end{figure}

Figure~\ref{fig:DeltaT} shows the distribution of total delay times between the coalescence of detected MBH binaries and the merger of their progenitor galaxies. Unsurprisingly, the inclusion of delay times results in significantly longer timescales between galaxy and MBH mergers, similar to what has been predicted from cosmological simulations \citep{Tremmel2018, Volonteri2020}. 
This fact has critical implications for multi-messenger astrophysics, as MBHs are likely to coalesce long after their host galaxies have merged. It is unlikely that by observing a clearly disturbed or an actively merging galaxy pair, one could identify the cradle of a LISA MBH binary (unless the MBH binary formed as a result of an even earlier galaxy merger event).
Electromagnetic counterparts to MBH binaries may therefore be commonly found in relatively undisturbed galaxies, billions of years following the progenitor galaxy merger event.

The timescale for LS models is also affected by the presence of SN feedback. As discussed previously, SN winds remove gas in low-mass, high-redshift galaxies and
therefore curtail MBH growth and gas-driven migration in those systems. 
This has a clear impact on Fig.~\ref{fig:DeltaT} (though not as dramatic as the inclusion of intermediate-scale delays). Indeed, by ejecting gas from
the nuclear region, SN explosions result in the suppression
of circumbinary disk migration, thereby increasing the typical delays between
galaxy and MBH coalescences.

\section{Quasar luminosity functions}\label{LF}

While we have discussed the predictions and implications of the model as they relate to gravitational-wave detection, we can also examine the model results against electromagnetic observations, namely the quasar luminosity function. We adopt the most recent estimate of the bolometric luminosity function \citep{2020arXiv200102696S}, and supplement it with an upper limit to the faint end at $z=6$. The latter is determined in X-rays \citep{2016MNRAS.463..348V}, but is transformed into bolometric luminosity, after applying a correction for Compton-thick AGNs \citep{2014ApJ...786..104U}, by using the same bolometric correction as in \cite{2020arXiv200102696S}.  

\begin{figure}
\centering
\includegraphics[width=0.45\textwidth]{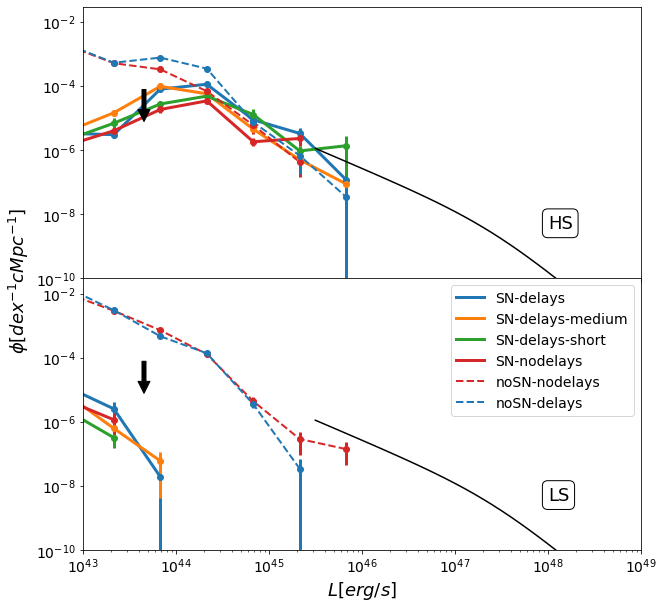}
\caption{Bolometric quasar luminosity function at $z=6$ for HS and LS models (colored lines, upper and lower panels, respectively). Black line: observed bolometric luminosity function \citep{2020arXiv200102696S}. Black arrows: upper limits on the X-ray luminosity function \citep{2016MNRAS.463..348V}, transformed into bolometric luminosity. }
\label{fig:Lum_bol_z6}
\end{figure}

All of our models (both LS and HS, with and without delays, with and without SN feedback) reproduce equally well the quasar luminosity function at $z<3$, with differences appearing only at higher redshift. We therefore focus our comparison  at $z=6$ (Fig.~\ref{fig:Lum_bol_z6}) and discuss lower redshifts only briefly. 

Dynamical delays do not have a noticeable effect on the luminosity function. Inclusion of the effect of SN feedback on MBH growth has the strongest impact on LS models. The specific implementation used here for SN feedback causes a clear underestimate of the luminosity function at $z=6$ for all LS models that include this effect.\footnote{Note that \citet{Habouzit2017} find relatively good agreement with the same upper limit starting with MBH seeds with mass $\sim 10^2-10^3 M_\odot$. This confirms  again that the specific implementation of how SN feedback affects MBH growth has a significant bearing on the results.} LS models without SN feedback also struggle somewhat to produce sufficiently massive MBHs and bright quasars, but overestimate the upper limit at the faint end. In summary, LS models without SN feedback overestimate the number of faint quasars, while LS models with SN feedback underestimate the number of bright AGNs. 

All HS models fare well with bright AGNs (note that the semi-analytic model does not include halos with mass $>10^{12}-10^{13}$ at $z=6$, where quasars above the knee of the luminosity function are expected to reside). HS models without SN feedback overestimate the faint end of the luminosity function, being above the upper limit derived by \cite{2016MNRAS.463..348V}. HS models with SN feedback are compatible with the upper limit at the faint end, and sit nicely on the observed portion of the luminosity function. 

An interesting point is that models with very different merger rates produce very similar luminosity functions. Table \ref{tab:rates_new_names} shows that the HS models \emph{SN-delays} and \emph{SN-nodelays} have a merger rate that differs by almost two orders of magnitude (25 versus 1269), but the luminosity functions are very similar. 
This comes about because  models
featuring the same (large) number of MBH seeds at high redshift, but with different merger rates due to dynamical delays,
can produce  luminosity functions  compatible with observations
if SN feedback efficiently suppresses  MBH accretion.

The results are similar for the various models at $z=5$, while by $z=4$ they start to converge to the same luminosity function, with LS and HS models faring equally well for bright quasars, and with only LS models without SN feedback slightly over-predicting the faint end down to $z=3$. At even lower redshift, all models produce indistinguishable luminosity functions. 

\section{Stochastic background in the pulsar-timing array band}\label{sec:stoch}

Observations of the unresolved stochastic background of gravitational waves by pulsar-timing array experiments are scientifically complementary
to LISA operations. These experiments include the European Pulsar Timing Array (EPTA) collaboration~\citep{Desvignes2016}; the Australian Parkes Pulsar Timing Array (PPTA) experiment~\citep{Reardon2016}; and the American NANOGrav collaboration~\citep{NANOGrav2015}. These experiments also share their data under the patronage of the International Pulsar Timing Array (IPTA) collaboration~\citep{ipta1,Verbiest2016,Perera2019}.

Pulsar-timing arrays attempt to detect gravitational waves by cross-correlating the timing residuals of ms pulsars~\citep{Sazhin1978,Foster1990}. The presence of a gravitational-wave stochastic background at frequencies $\sim $ nHz would produce a potentially detectable quadrupolar correlation between the residuals of pulsars at different sky locations~\citep{Hellings1983}. The same technique can also detect individual gravitational-wave signals, if those are strong enough to be resolved above the stochastic background~\citep{Sesana2009,Babak:2015lua,nanograv11resolved}.

The stochastic background at nHz frequencies is expected to be produced mainly by inspiraling binaries of MBHs with total masses between $10^8$ and $10^{10} \rm ~M_\odot$ at redshifts $\lesssim 2$~\citep{Wyithe2003,Sesana_Vecchio2008,McWilliams2014,Rajagopal1995,Jaffe2003,Sesana2013,Ravi2015,Sesana2016,Sesana2009,Ravi2012,Kulier2015,Kelley2017,Bonetti2018b}. While these systems are heavier than those targeted by LISA, their hierarchical formation mechanism is the same (halo/galaxy mergers followed by the formation of MBH pairs/binaries). Therefore, pulsar-timing arrays provide a distinct, complementary way of exploring astrophysics similar to LISA's, but at larger scales and in different host environments.

Interestingly, the aforementioned pulsar-timing array experiments are already ongoing and analyzing data, and have put strong upper bounds on the stochastic background in the nHz band~\citep{Lentati2015,Arzoumanian2016,Shannon2015,Verbiest2016,Perera2019}. The most robust constraint to date comes from NANOGrav's 11-yr data set~\citep{Arzoumanian2018}. Stronger bounds have been put forward by PPTA~\citep{Shannon2015}, but it is unclear if those robustly account for uncertainties in the position of the solar system barycenter. Overall, these upper bounds have  already put constraints on models of MBH mergers~\citep{Wyithe2003,Sesana_Vecchio2008,McWilliams2014,Rajagopal1995,Jaffe2003,Sesana2013,Ravi2015,Sesana2016,Sesana2009,Ravi2012,Kulier2015,Kelley2017,Bonetti2018b,2018NatCo...9..573M}, even ruling out the most extreme ones~\citep{McWilliams2014} in which such mergers are very abundant. 

\begin{figure}
\centering
\includegraphics[width=0.45\textwidth]{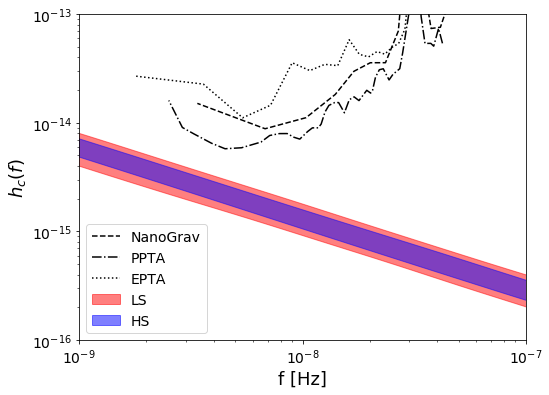}
\caption{Characteristic strain of the stochastic background from MBH binaries in the band of pulsar-timing array experiments. The red and blue shaded regions encompass the LS and HS models, respectively. Unlike for the merger rate in the LISA band, the predictions for the pulsar-timing array signal are quite robust and show only a mild dependence on the model. Also shown are the sensitivity curves of ongoing   pulsar-timing array experiments~\citep{Arzoumanian2018,Shannon2015,Desvignes2016}.}
\label{fig:MBH_all_models}
\end{figure}

In Fig.~\ref{fig:MBH_all_models}, we show the predictions of the models presented in this paper for the background's characteristic strain $h_c$, as a function of gravitational-wave frequency and compared to the upper  bounds from EPTA, PPTA and NANOGrav  \citep[see e.g.][for the definition and computation of $h_c$]{Phinney2001,Sesana_Vecchio2008,Dvorkin2017}. The pink and purple shaded areas denote the envelope of the predictions of  all models
(i.e. the minimum area enclosing  the predictions of all models)
for our LS and HS scenarios, respectively. 
The slope of the background follows from the assumption that circular binaries lose energy only through the emission of gravitational waves.
Indeed, most of the background signal comes from MBH binaries in their early inspiral phase, where GW emission is well described by the quadrupole formula,
which gives $h_c\propto f^{-2/3}$. Note that the predictions from all models are very similar, as the signal is mostly emitted by binaries involving MBHs with masses above $10^8  M_\odot$ at low redshift. For these systems, the impact of the different seeding and delay prescriptions is minor.

Reassuringly, the scatter of the predictions of the different models is very small, unlike what happens for the LISA detection rate predictions shown above. This  was to be expected~\citep[c.f. e.g.][]{Bonetti2018b}, because the dependence on the seeding mechanism fades out when MBHs evolve to very large masses, and because
the bulk of the pulsar-timing array signal comes from comparable-mass MBH binaries, for which the delays between galaxy and MBH merger are generally shorter~\citep{Dvorkin2017}. By comparing to the results of \citet{Bonetti2018b}, Fig.~\ref{fig:MBH_all_models} suggests that the stochastic background may be detected by pulsar-timing arrays in $\approx 15-20$ yr of data collection, assuming a putative array of 50 ms pulsars monitored at the 100 ns level of precision. A significantly earlier detection would  be achievable with
the Square Kilometer Array (SKA) telescope~\cite{Dvorkin2017}.

\section{Conclusions}\label{sec:conclusions}

In this paper, we explore how the expected event rates
for MBH mergers and their corresponding gravitational wave signals
depend on the physical processes delaying the evolution of MBHs
at intermediate separations ($\sim 10s - 1000s$ pc) and on processes like SN feedback, which can regulate  their growth by accretion and affect their gas-driven migration.
To this purpose, we perform semi-analytic simulations of the co-evolution of galaxies and MBHs, starting from either low- or high-mass seeds for the MBH population at high redshift. We find that, regardless of the MBH seed mass model, the predicted rates of MBH mergers are heavily dependent on the inclusion of both delayed binary formation and SN-regulated MBH growth.

The effect of including even moderate delay timescales (i.e. the inclusion of dynamical friction timescales on scales below 100 pc as in our delays-short models) results in significantly fewer MBH mergers for models incorporating high-mass MBH seeds. Merger rates from low-mass seed models are less sensitive to such delay timescales when SN-regulated growth is included. The predicted MBH merger rates are also relatively insensitive to the detailed model choices for binary formation timescales. However, the inclusion of galactic-scale dynamical evolution (i.e. from several kpc to $\sim100$ pc separations), a phase of evolution often disregarded in semi-analytic models, results in an additional factor of $\sim2-4$ decrease in MBH merger and detection rates for both low and high-mass seed models. The inclusion of these delay timescales mostly affects the abundance of high-redshift, low-mass ($M_{\rm tot} < 10^6 M_{\odot}$) mergers in the case of high-mass MBH seeds. For models assuming low-mass MBH seeds, binary formation delay timescales also affect the tail of the mass ratio distribution of MBH mergers.

Feedback from SN winds primarily affects the merger rate of low-mass MBH seeds. In these models, SN feedback is able to expel gas from the center of low-mass galaxies/dark matter halos, thereby regulating the gas reservoir available to MBHs in the early universe. For low-mass seed models, on the one hand this decreases the intrinsic merger rate of MBHs, as the depletion of gas results in longer binary hardening timescales. On the other hand, SN winds also suppress seed growth via accretion, preventing MBHs from growing to sufficiently high masses for their mergers to be detectable with LISA.

Models with high-mass MBH seeds are generally less affected by SN feedback due to their already substantial masses, which typically are already detectable by LISA. However, when binary formation delay times are included, SN feedback results in slightly more numerous MBH mergers. This happens because rapid, early mass growth spins MBHs up if SN feedback is unaccounted for. As a result, when MBHs merge at high redshift, the MBH remnant forming from the merger experiences a strong gravitational recoil kick. The kick can even be high enough to remove the MBH remnant from the host galaxy, depleting the number of central MBHs available to experience future mergers.

Another important consequence of SN-regulated MBH growth is on the predicted luminosity function at high redshift. Both low- and high-mass seed models predict too many low-luminosity MBHs without SN feedback. The effect of SN feedback is less dramatic for high-mass MBH seeds, but still significant. In summary, only models that include SN-regulated MBH growth are consistent with high-$z$ quasar observations.

The most pessimistic model that we use in this work, which includes the longest binary formation delay times as well as SN-regulated growth, predicts that LISA should be able to detect several MBH mergers during its nominal mission duration of 4 yr. This work highlights how LISA will be a critical tool for constraining and discriminating models of MBH growth and dynamical evolution. Conversely, we predict that pulsar-timing array detection of the gravitational-wave stochastic background should be relatively insensitive to model variations, including, unsurprisingly, the MBH seed masses. Rather, our results robustly show that the stochastic background should be detectable in the near future as the sensitivity of pulsar-timing arrays improves~\citep{Bonetti2018b,Dvorkin2017}, and that this remains true even for dramatically different MBH evolutionary models.

\acknowledgments

We acknowledge the financial support provided under the European Union's H2020 ERC Consolidator Grant ``GRavity from Astrophysical to Microscopic Scales'' grant agreement no. GRAMS-815673. I.D. thanks SISSA for hospitality during the early stages of this work. This work has made use of the Horizon Cluster, hosted by the Institut d'Astrophysique de Paris. We thank Stephane Rouberol for  smoothly running this cluster for us.

\bibliography{biblio}{}
\bibliographystyle{aasjournal}



\end{document}